\RequirePackage{textcomp,graphicx,multirow}
\documentclass[%
reprint,
showkeys,
amsmath,amssymb,
aps,
prb,
floatfix,
]{revtex4-1}
\usepackage[utf8]{inputenc}
\usepackage[english]{babel}
\usepackage{graphicx}
\usepackage{amsmath}
\usepackage{amssymb}
\usepackage{natbib}

\usepackage{color,textcomp,multirow}

\begin{document}
	
\title{Surface pinning and triggered unwinding of skyrmions in a cubic chiral magnet}
	
\author{Peter~Milde}
\email{peter.milde@tu-dresden.de}

\author{Erik~Neuber}
\affiliation{Institute of Applied Physics, Technische Universit\"at Dresden, D-01062 Dresden, Germany}

\author{Andreas~Bauer}
\author{Christian~Pfleiderer}
\affiliation{Physik-Department, Technische Universit\"{a}t M\"{u}nchen, D-85748 Garching, Germany}

\author{Lukas~M.~Eng}
\affiliation{Institute of Applied Physics, Technische Universit\"at Dresden, D-01062 Dresden, Germany}
\affiliation{Center for Advancing Electronics Dresden (cfaed), Technische Universit\"at Dresden, D-01062 Dresden, Germany}

\date{\today}
	
\begin{abstract}
	In the cubic chiral magnet Fe$_{1-x}$Co$_{x}$Si a metastable state comprising of topologically nontrivial spin whirls, so-called skyrmions, may be preserved down to low temperatures by means of field cooling the sample. This metastable skyrmion state is energetically separated from the topologically trivial ground state by a considerable potential barrier, a phenomenon also referred to as topological protection. Using magnetic force microscopy on the surface of a bulk crystal, we show that certain positions are preferentially and reproducibly decorated with metastable skyrmions, indicating that surface pinning plays a crucial role. Increasing the magnetic field allows an increasing number of skyrmions to overcome the potential barrier, and hence to transform into the ground state. Most notably, we find that the unwinding of individual skyrmions may be triggered by the magnetic tip itself, however, only when its magnetization is aligned parallel to the external field. This implies that the stray field of the tip is key for locally overcoming the topological protection. Both the control of the position of topologically nontrivial states as well as their creation and annihilation on demand pose important challenges in the context of potential skyrmionic applications.
\end{abstract}
	

\maketitle

\section{Introduction}

A certain type of spin whirls, so-called skyrmions, have become a vividly investigated topic in the field of solid state research for the last decade. The key characteristic of these objects is their nontrivial topology, i.e., skyrmions may not be continuously transformed into a topologically trivial state, such as a ferromagnet or helimagnet. As a consequence, typical skyrmion states are energetically separated from topologically trivial states by a considerable potential barrier and, once formed, skyrmions tend to be remarkably robust. This robustness together with a plethora of static and dynamic properties intimately connected with the nontrivial topology such as emergent electrodynamics\cite{2010_Jonietz_Science, 2012_Schulz_NatPhys, 2012_Yu_NatCommun, 2013_Fert_NatNanotechnol} or characteristic collective microwave modes\cite{2012_Mochizuki_PhysRevLett, 2012_Onose_PhysRevLett, 2015_Schwarze_NatMater, 2015_Okamura_PhysRevLett, 2016_Seki_PhysRevB} promises novel concepts for spintronic device applications.\cite{2013_Nagaosa_NatNanotechnol, 2017_Fert_NatRevMater} In this context, the controlled creation and annihilation of single skyrmions remains an important task.

Over the years, magnetic skyrmions have been identified in a wide range of materials including monolayers\cite{2011_Heinze_NatPhys, 2016_Wiesendanger_NatRevMater}, thin films\cite{2015_Jiang_Science, 2016_Boulle_NatNanotechnol, 2016_Woo_NatMater, 2018_Woo_NatCommun}, and heterostructures\cite{2016_Moreau-Luchaire_NatNanotechnol} as well as bulk compounds\cite{2009_Muhlbauer_Science, 2015_Kezsmarki_NatMater, 2017_Kurumaji_PhysRevLett, 2017_Nayak_Nature}. All these systems share a lack of inversion symmetry, either introduced by the surface or an interface, or due to a non-centrosymmetric crystal structure, enabling chiral Dzyaloshinsky--Moriya spin--orbit interactions to twist the magnetic structure on length scales much larger than atomic distances. In this respect, the perhaps most intensively studied class of compounds are the bulk cubic chiral magnets, notably MnSi\cite{2009_Muhlbauer_Science, 2012_Tonomura_NanoLett}, Fe$_{1-x}$Co$_{x}$Si\cite{2010_Munzer_PhysRevB, 2010_Yu_Nature}, FeGe\cite{2011_Yu_NatMater, 2013_Moskvin_PhysRevLett}, and Cu$_{2}$OSeO$_{3}$\cite{2012_Seki_Science, 2012_Adams_PhysRevLett} crystallizing in space group $P2_{1}3$, as well as $\beta$-Mn-type Co-Zn-Mn alloys\cite{2015_Tokunaga_NatCommun} crystallizing in space group $P4_{1}32$ or $P4_{3}32$. The magnetic phase diagrams of these compounds are highly reminiscent of each other, despite distinctly different characteristic length, temperature, and magnetic field scales.\cite{2016_Bauer_Book} In particular, in a phase pocket just below the ordering temperature $T_{c}$ at finite magnetic field a trigonal lattice of skyrmions is observed in the plane perpendicular to that field, forming skyrmion tubes along the field direction.

The topological protection of these skyrmions is reflected in the possibility to metastably extend the temperature and field range for which they are observed by means of field cooling through the reversible phase pocket. In undoped systems, such as MnSi, FeGe, or Cu$_{2}$OSeO$_{3}$, rapid quenching\cite{2016_Oike_NatPhys, 2017_Nakajima_SciAdv, 2018_Berruto_PhysRevLett} or additional tuning by means of pressure\cite{2013_Ritz_PhysRevB} or electrical field\cite{2016_Okamura_NatCommun} is required. In contrast, the intrinsically slower time scales due to disorder and defects in doped systems, such as Fe$_{1-x}$Co$_{x}$Si or $\beta$-Mn-type Co-Zn-Mn alloys, allow to obtain a metastable skyrmion lattice state already at moderate cooling rates.\cite{2010_Munzer_PhysRevB, 2016_Bauer_PhysRevB, 2016_Karube_NatMater, 2017_Karube_PhysRevMater, 2018_Bauer_Book} At low temperatures the dynamics of the magnetic system are completely frozen-in and the thermal energy is no longer sufficient to overcome the potential barrier that separates the skyrmion state from the topologically trivial helimagnetic states. In turn, adjusting the energy landscape by changing the applied magnetic field permits to study the topological unwinding processes and their dynamics directly in real space using techniques such as Lorentz transmission electron microscopy (LTEM) and magnetic force microscopy (MFM).

Although the experimental observation of single skyrmions in a topologically trivial magnetic background was already reported in the seminal study by Yu et al. using LTEM on a thin bulk sample of Fe$_{0.5}$Co$_{0.5}$Si,\cite{2010_Yu_Nature} the properties of single skyrmions were so far mostly studied in heterostructures and thin-film systems. Notable examples include Pd/Fe bilayers on Ir(111) in which individual skyrmions pinned at defects could be created and destroyed by injecting local spin-polarized currents from the tip of a scanning tunneling microscope\cite{2013_Romming_Science} and detected by means of tunneling non-collinear magnetoresistance\cite{2015_Hanneken_NatNanotechnol}. In the same spirit, individual skyrmions were reversibly created and destroyed by means of local electric fields in trilayers of Fe on Ir(111).\cite{2017_Hsu_NatNanotechnol} 
In multilayers based on stacks of Pt and 3d transition-metal ferromagnets, single skyrmions were created and manipulated exploiting the local magnetic field of a MFM tip \cite{2019_Casiraghi_arXiv, 2018_Zhang_APL}.
Furthermore, the presence of a single skyrmion could be detected electrically via the anomalous Hall effect\cite{2018_Maccariello_NatNanotechnol} while current-driven dynamics were investigated by means of x-ray microscopy.\cite{2015_Buttner_NatPhys, 2016_Woo_NatMater, 2017_Litzius_NatPhys} 

In contrast, in bulk chiral magnets real-space studies on the solitonic character as well as on the creation, destruction, stability, and dynamics of skyrmions typically addressed large ensembles,\cite{2013_Milde_Science, 2014_Mochizuki_NatMater, 2015_Rajeswari_ProcNatlAcadSciUSA, 2017_Pollath_PhysRevLett, 2018_Berruto_PhysRevLett} focusing on the fundamental mechanisms. For instance, using MFM and LTEM on Fe$_{0.5}$Co$_{0.5}$Si it was established that for decreasing and increasing magnetic fields skyrmions unwind by two distinctly different mechanisms, notably the coalescence of neighboring skyrmions\cite{2013_Milde_Science} and the pinching-off of skyrmion tubes\cite{2017_Wild_SciAdv}, respectively. In this context, MFM provides an extremely viable tool as it not only allows to map the magnetic texture across a large area in real space, but also allows to manipulate the magnetic system by means of carefully exploiting the interplay of the magnetic tip with the sample\cite{2017_Wang_NJP}. Note in this context that additional low-temperature skyrmion phases were recently observed in cubic chiral magnets in the presence of sufficiently strong magnetocrystalline anisotropies\cite{2018_Chacon_NatPhys} or magnetic frustration\cite{2018_Karube_SciAdv}. While thermal fluctuations are key for the thermodynamic stabilization of the high-temperature phase pocket,\cite{2009_Muhlbauer_Science, 2013_Buhrandt_PhysRevB} these findings highlight that in fact skyrmions may be realized along different routes---even within one and the same material---in turn permitting novel concepts for the realization, manipulation, and exploitation of topologically non-trivial spin structures.

In this communication, we use MFM to investigate the metastable skyrmion lattice on the surface of a bulk single crystal of Fe$_{0.5}$Co$_{0.5}$Si. We show that pinning sites, most likely at the surface of the bulk sample, reproducibly confine single skyrmions to preferential positions. When increasing the magnetic field at low temperatures, isolated skyrmions tend to survive at exactly these positions. Perhaps most notably, however, we find that the unwinding of single skyrmions may be triggered by the magnetic tip when its magnetization is chosen such that its stray field locally enhances the external magnetic field. These results demonstrate that bulk chiral magnets may be exploited as hosts for controllable, topologically non-trivial entities.

\section{Experiment}

\begin{figure}[ht]
	\includegraphics[width=\columnwidth]{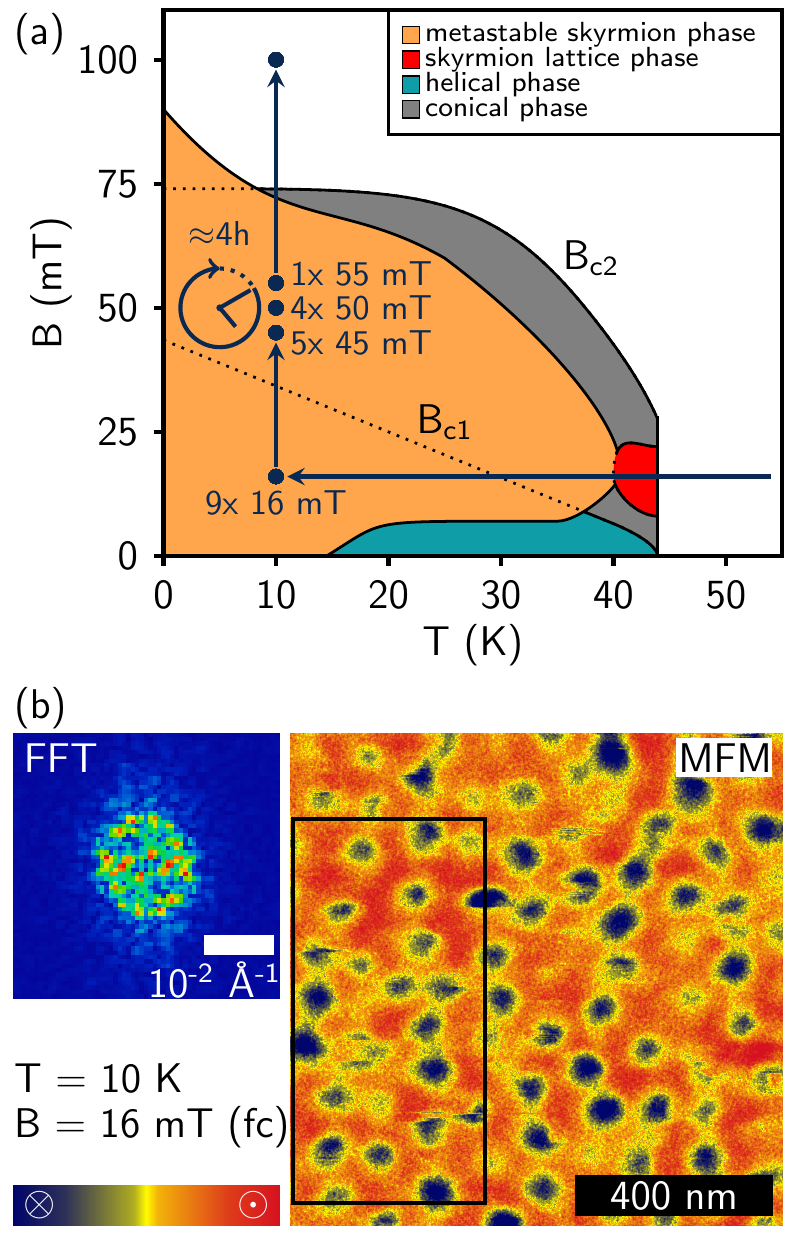}
	\caption{Magnetic phase diagram (a) and typical magnetic force microscopy data (b) of Fe$_{0.5}$Co$_{0.5}$Si. (a)~Schematic magnetic phase diagram as obtained by means of field cooling through the thermodynamically stable pocket of skyrmion lattice (red shading). When subsequently increasing or decreasing the magnetic field, a metastable skyrmion state (orange shading) survives over a large regime in field and temperature. Dotted lines indicate phase boundaries as observed after zero-field cooling. (b)~Typical magnetic force microscopy data as measured after field cooling the skyrmion lattice down to 10~K. The black rectangular marks the area of interest for the analysis described in the following. Note the considerable amount of disorder in the skyrmion state, reflected also in the Fast Fourier Transform shown on the left-hand side.}
	\label{figure1}
\end{figure}

In the present study, the carefully polished surface of a bulk single crystal of Fe$_{0.5}$Co$_{0.5}$Si of $2.7\times1.7\times0.5~\mathrm{mm}^{3}$ was investigated. Using a wire saw, the sample was prepared from a large single-crystal ingot grown by means of the optical floating-zone technique.\cite{2011_Neubauer_RevSciInstrum, 2016_Bauer_RevSciInstrum, 2016_Bauer_PhysRevB} The largest surface was aligned perpendicular to a crystalline $\langle110\rangle$ axis and was mechanically polished. The magnetic phase diagram of the specimen, depicted in Figure.~\ref{figure1}(a), was inferred from measurements of the magnetization and the magnetic ac susceptibility prior to the MFM study, cf. reference \citenum{2018_Bauer_Book}, and may be described as follows: Starting from the paramagnetic state at high temperatures, helimagnetic order with a wavelength of $\lambda = 90$~nm emerges below the transition temperature $T_{c} = 45$~K.\cite{1983_Beille_SolidStateCommun, 2007_Grigoriev_PhysRevB} When cooling down in zero field, the helical state is observed, which is characterized by multiple macroscopic domains of helices propagating along one of the easy $\langle100\rangle$ axes, implied by the weak magneto-crystalline anisotropies.\cite{1980_Bak_JPhysC, 1986_Ishimoto_JMagnMagnMater, 2006_Uchida_Science} Increasing the magnetic field at low temperatures results in the re-orientation of the propagation vector into the field direction at $B_{c1}$, giving rise to the single-domain conical state. In the latter, the spins increasingly tilt towards the field direction with increasing field until a field-polarized state is reached at the upper critical field $B_{c2}$. Just below $T_{c}$ at intermediate magnetic fields, a single pocket of skyrmion lattice state is observed. When field cooling through this phase pocket, at an applied field of 16~mT in the present case, the skyrmion lattice state is metastably preserved already at cooling rates of the order of K/min.\cite{2010_Munzer_PhysRevB, 2016_Bauer_PhysRevB} Following this field cooling process, at low temperature and constant field dynamics are frozen-in and the magnetic texture remains qualitatively unchanged, at least up to the time scale of typical experiments (several hours). Increasing or decreasing the magnetic field ultimately results in the unwinding of the metastable skyrmions into the respective ground state.\cite{2018_Bauer_Book}

Magnetic force microscopy was performed on an Omicron low-temperature ultra-high vacuum atomic force microscope equipped with RHK Technology Inc.\ R9-controller electronics using Nanosensor SSS-QMFMR probes. In all MFM measurements, the contact potential between cantilever and sample was compensated by running a Kelvin probe force controller -- see appendix \ref{Appendix:KPFM} for details. The magnetization direction of the tip was fixed by application of a large magnetic field along the positive $z$ direction. Consequently, in the images present here, blue and red color corresponds to magnetization components pointing into and out of the plane of sight, respectively. These images were recorded in a two-step process. First, the topography of the sample was measured along the edges of the scan area only, thereby avoiding any disturbance of the magnetic state, and the two-dimensional slope was canceled. Second, the MFM tip was retracted by ${\sim}20$~nm in order to record the magnetic forces while scanning the sample surface in the area of interest. In the MFM images, the slow scan direction was from top to bottom ($y$ direction), while fast scans were performed along the horizontal $x$ direction. Scans from left to right and from right to left may be treated separately. Note that recording an MFM image across an area of $2 \times 2~\mu\mathrm{m}^{2}$ ($512 \times 512$~pixels, 1~s/line) takes ${\sim}17$~minutes.

Fixing the sample position, we repeatedly field cooled the sample in an applied field of 16~mT from well above $T_{c}$ to $T = 10$~K at an average rate of 0.4~K/min. At the target temperature, an MFM image of the initial magnetic texture was recorded. Subsequently, the magnetic field was increased to one of the target values, namely five times to 45~mT, three times to 50~mT, and once to 55~mT. After the target field was reached, the recording of a MFM image was started, followed by consecutive images every 17~minutes for about four hours. Thereafter, finally, the field was increased to 100~mT and an image for background correction was measured in the field-polarized state. Note that the MFM tip is retracted to a safe position during the field cooling process and repositioned for the actual measurement. In order to compensate the small horizontal drift between separate cooling cycles, topographic signatures are used to determine the exact position on the sample.

The analysis of the MFM data was carried out by means of the Pygwy interface of the Gwyddion analysis software \cite{2012_Necas_CentEurJPhys} in an iterative multi-step process. Details are given in appendix \ref{Appendix:Data}.

\section{Results}

\subsection{Surface pinning}

A typical example of an MFM image recorded after field cooling is depicted in Figure~\ref{figure1}b. The applied magnetic field is pointing out of the plane of sight. In contrast to the well-ordered trigonal skyrmion lattice state observed for instance in the reversible phase pocket of Cu$_{2}$OSeO$_{3}$\cite{2016_Milde_NanoLett}, the metastable skyrmion state as observed on the surface of bulk Fe$_{0.5}$Co$_{0.5}$Si exhibits considerable amounts of disorder, somehow reminiscent of a frozen-in glassy state. Consequently, the corresponding Fast Fourier Transform lacks the clear sixfold symmetry. Still, the majority of the skyrmions possesses a well-defined essentially circular shape. Some skyrmionic objects, however, seem distinctly deformed or split in parts. We attribute the splitting to partly reversible changes of the position of single skyrmions between two neighboring pinning sites that are triggered by the stray field of the magnetic tip while scanning the image from top to bottom. Such tip-induced processes will be discussed in further detail below.

\begin{figure}
\includegraphics[width=\columnwidth]{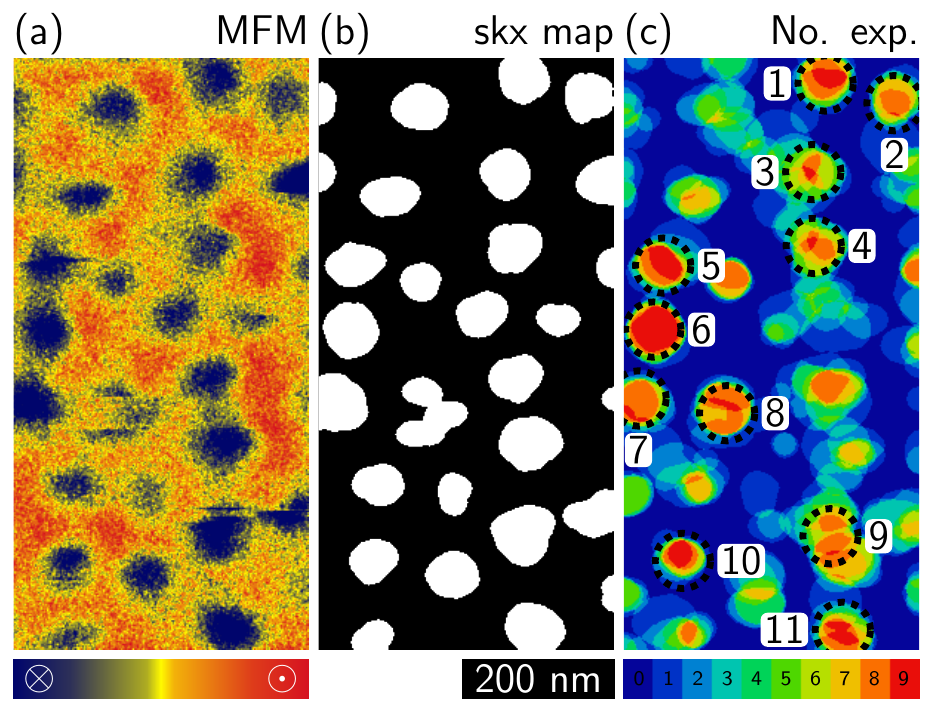}
\caption{Typical initial magnetic texture after field cooling the sample in an applied field of 16~mT down to 10~K. (a)~Magnetic force microscopy image. Blue and red color corresponds to magnetization components parallel and anti-parallel to the line of sight. (b)~Black-and-white map of the initial magnetic texture inferred by thresholding the data of panel (a). White objects correspond to skyrmions. (c)~Color map obtained by summing up the black-and-white maps of the nine independent field cooling cycles. Red color indicates positions where a skyrmion was observed after each field cooling.}
\label{figure2}
\end{figure}

In the following, we concentrate on a region of $0.4 \times 0.8~\mu\mathrm{m}^{2}$, marked by the black rectangle in Figure~\ref{figure1}, that was covered in all measurement runs. An enlarged view of this area of the sample is depicted in Figure~\ref{figure2}a. Thresholding the MFM images allows to identify the positions of individual skyrmions in form of black-and-white maps, as shown in Figure~\ref{figure2}b. Summing up these maps, hence comparing the initial magnetic textures for the different field cooling runs, we identify certain positions that are reproducibly decorated by a skyrmion. In the color map in Figure~\ref{figure2}c the corresponding positions are shown in red color, marked by dashed circles, and numbered. The attractive potential for skyrmions at these positions may be attributed to defect-related pinning, consistent with the results of micromagnetic simulations and calculations using density functional theory.\cite{2015_Muller_PhysRevB, 2016_Choi_PhysRevB} 
Taking into account that metastable skyrmion strings in bulk samples are typically segmented \cite{2017_Kagawa_NatComm} and of finite stiffness \cite{2019_Kravchuk_arXiv}, these defect sites probably are located at or close to the surface of the sample. No information on the behavior deep inside the bulk is available. When trying to understand the latter microscopically, potential surface twisting of the magnetic texture needs to be taken into account, as recently observed in terms of the N\'{e}el-type twisting of the skyrmion spin texture at the surface of Cu$_{2}$OSeO$_{3}$ by means of an analysis of the polarization dependence of resonant elastic x-ray scattering.\cite{2018_Zhang_ProcNatlAcadSciUSA} Such an in-depth description, however, is beyond the scope of the present study.

\begin{figure}
\includegraphics[width=\columnwidth]{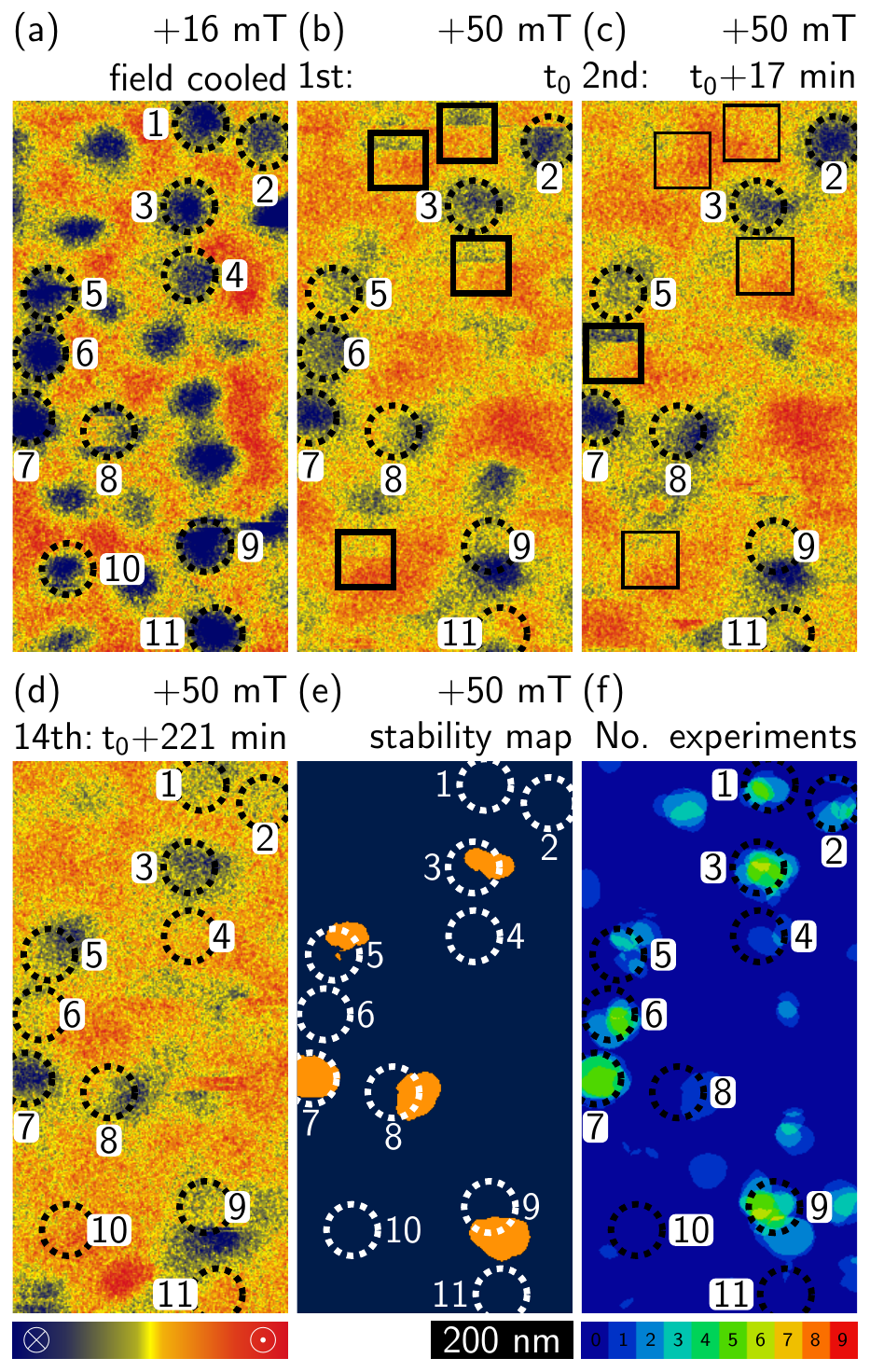}
\caption{Typical magnetic force data as measured after field cooling and subsequently increasing the magnetic field. (a)~Initial magnetic texture after field cooling in an applied field of 16~mT down to 10~K. (b)~Magnetic texture after increasing the magnetic field to 50~mT. Black squares mark incidences where skyrmions disappears while being scanned. (c)~Subsequent MFM image, started 17~minutes after reaching 50~mT. (d)~Final MFM image of the series, started 221~minutes after reaching 50~mT. (e)~Blue-and-orange map where white color indicates positions for which a skyrmion is observed throughout the whole series of MFM images, see text for details. (f)~Color map obtained by summing up the black-and-white maps of the nine independent field cooling cycles. Light green color indicates positions where a skyrmion was observed during most experiments.}
\label{figure3}
\end{figure}

Next, in Figure~\ref{figure3} we analyze how these spots of attractive potential for individual skyrmions affect the stability under increasing magnetic field. The initial spin texture after field cooling is shown in Figure~\ref{figure3}a, with dashed circles marking the positions that are reproducibly decorated with an individual skyrmion after each field cooling as described above. The MFM image recorded directly after increasing the magnetic field, in this case to 50~mT, is shown in Figure~\ref{figure3}b. Changing the applied field results in an altered energy landscape thereby driving the unwinding of the metastable skyrmion state towards the topologically trivial ground state. Consequently, several skyrmions have vanished already compared with the initial distribution. Keeping temperature and magnetic field unchanged, consecutive MFM images are recorded. Figures~\ref{figure3}c and \ref{figure3}d depict the second MFM image, started 17~minutes after increasing the field, as well as the 14th and final image, started after 221~minutes. As time elapses further skyrmions disappear, consistent with previous reports. 
Note that, although the skyrmion lattice in chiral magnets is usually described in terms of continuous mean-field model, the fact that single skyrmions decay or survive, respectively, already highlights distinct solitonic character of the skyrmions in Fe$_{0.5}$Co$_{0.5}$Si.

\subsection{Skyrmion stability}

\begin{table}[b!]
	
		\begin{ruledtabular}
			\begin{tabular}{@{}c|c|c|c|cr|cr|cr@{}}
				field & \multicolumn{3}{c|}{No. of skyrmions} & \multicolumn{6}{c}{No. of skyrmions after 4h} \\ 
				& pinned & non-pin. & $\Sigma$  & \multicolumn{2}{c|}{pinned} & \multicolumn{2}{c|}{non-pin.} & \multicolumn{2}{c}{$\Sigma$}  \\ \hline
				\multirow{5}*{45 mT}	& 11 & 17 & 28 & 5 & (45\%) & 4 & (24 \%) & 9  & (32 \%) \\ \cline{2-10}
				& 11 & 21 & 32 & 5 & (45\%) & 1 & (5 \%)  & 6  & (19 \%) \\ \cline{2-10}
				& 10 & 15 & 25 & 7 & (70\%) & 2 & (13 \%) & 9  & (36 \%) \\ \cline{2-10}
				& 11 & 20 & 31 & 5 & (45\%) & 1 & (5 \%)  & 6  & (19 \%) \\ \cline{2-10}
				& 12 & 13 & 25 & 9 & (75\%) & 4 & (31 \%) & 13 & (52 \%) \\ \hline
				\multirow{3}*{50 mT}	& 11 & 17 & 28 & 4 & (36\%) & 2 & (12 \%) & 6  & (21 \%) \\ \cline{2-10}
				& 12 & 16 & 28 & 1 & (8\%)  & 1 & (6 \%)  & 2  & (7 \%)  \\ \cline{2-10}
				& 11 & 18 & 29 & 5 & (45\%) & 0 & (0 \%)  & 5  & (17 \%) \\ \hline
				55 mT & 11 & 17 & 28 & 2 & (18\%) & 1 & (6 \%)  & 3  & (11 \%) \\ \hline 
				average & \multicolumn{3}{c|}{ }  &   & 43 \%   &    & 11 \% & & 24 \% 
			\end{tabular}
		\end{ruledtabular}
		\caption{\label{table1} Comparison of the stability of pinned and non-pinned skyrmions as well as the number of skyrmions observed after the total waiting time of 4h. The survival probability of pinned skyrmions is roughly four times higher.}
\end{table}

Further analysis is carried out by means of thresholding each image of a decay series and multiplying the result. This procedure yields the blue-and-orange map shown in Figure~\ref{figure3}(e) indicating positions that are occupied by a skyrmion throughout the entire series of MFM images (orange color), i.e., skyrmions that are rather robust with respect to the increase of the magnetic field. 
Summing up the blue-and-orange maps of the nine independent field cooling cycles finally yields the color map shown in Figure~\ref{figure3}f. While at neither position a skyrmion is observed throughout all nine experiments, in particular positions that are reproducibly decorated by individual skyrmions after each field cooling process (dashed circles) also exhibit a high probability of still being occupied by a skyrmion after increasing the magnetic field and waiting for roughly 4~hours needed to acquire 14 MFM images.
A detailed comparison of the initial skyrmion positions, such as shown in Figure \ref{figure2}(b), with the as-obtained stability map allows to count the number of surviving pinned skyrmions and non-pinned skyrmions separately, summarized in table \ref{table1}. In average, we find that $\approx 43 \%$ of the pinned skyrmions but only $\approx 11 \%$ of the non-pinned skyrmions survive until the end of the image series,
i.e., the pinned skyrmions have a roughly four times higher survival probability. 
This finding implies that defect-related pinning not only fixes the position of a given skyrmion but also may effectively enlarge the local potential barrier preventing its unwinding, in turn enhancing robustness of the skyrmion against external perturbations, such as changes of the magnetic field or temperature. 
Note, however, that there are exceptions as some of the pinning sites in fact do not enhance the skyrmion stability, pointing towards a different nature of these sites, potentially related to topological defects within the skyrmion string \cite{2017_Kagawa_NatComm}, see for instance position No. 10 in Figure \ref{figure3}(f).

\subsection{Skyrmion collapse}

Perhaps most remarkably, in our experiments two types of unwinding processes may be distinguished. First, the signal associated with a given skyrmion completely vanishes from one MFM image to the next. This disappearance is attributed to the spontaneous unwinding of the skyrmion by virtue of a Bloch point during the time interval of ${\sim}17$~minutes between two consecutive MFM images. Second, the upper portion of the skyrmion is still visible while the lower portion is not, i.e., the signal disappears while the MFM tip scans across the given skyrmion. Corresponding incidents are marked by black squares. Further note that in all MFM images following thereafter, the respective skyrmion is missing. In principle, the unwinding of a given skyrmion may take place while the MFM tip is scanning across the given sample area just by coincidence. 
The large number of these events in our relatively small sampling, however, strongly favors a different interpretation, notably that the unwinding is triggered by the MFM tip
similar to reports of skyrmion manipulation with MFM tips in metal thin films \cite{2019_Casiraghi_arXiv,2018_Zhang_APL}. 

\begin{figure}
	\includegraphics[width=\columnwidth]{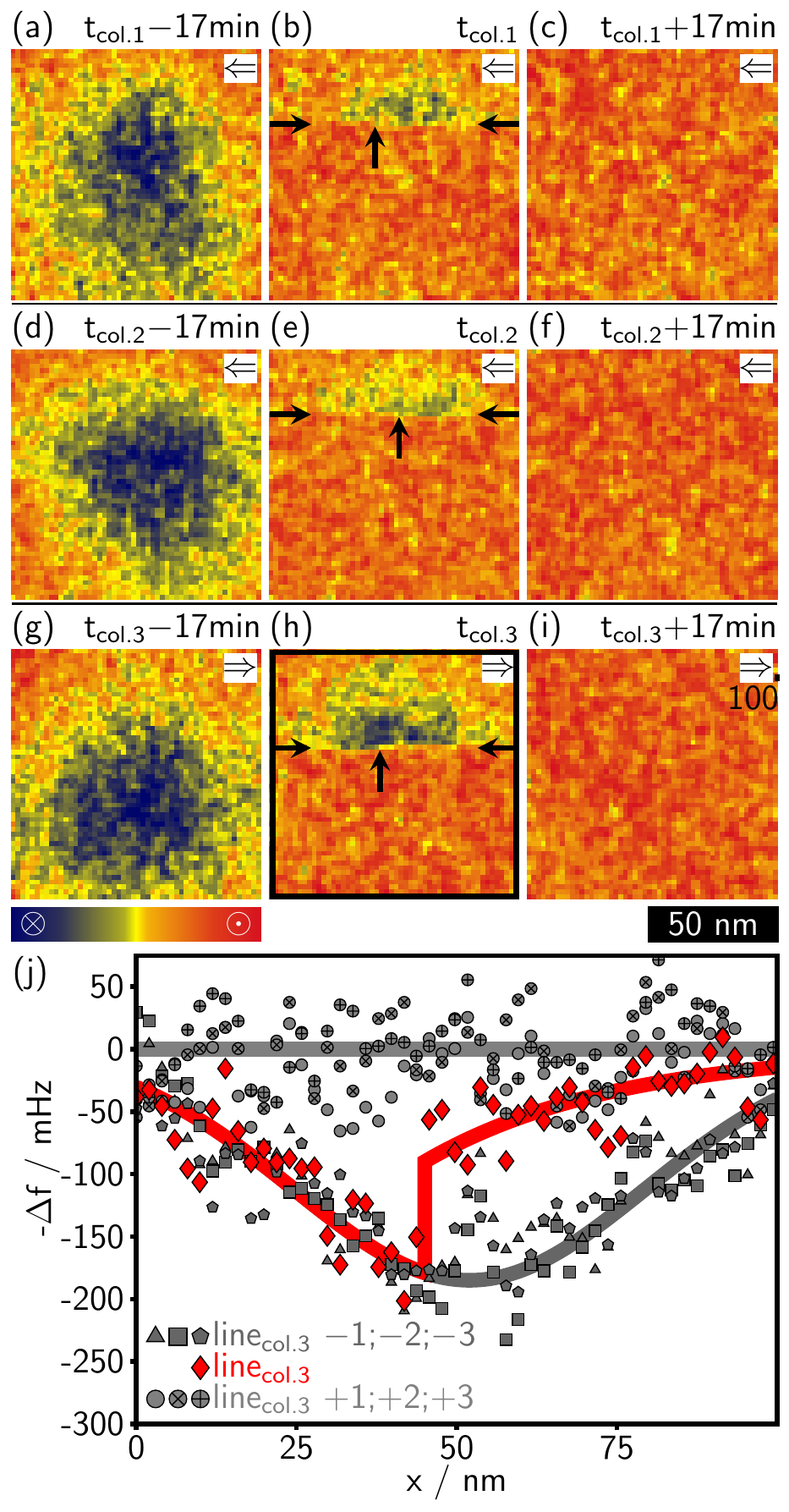}
	\caption{Magnetic force microscopy data of the triggered unwinding of single skyrmions. Three typical examples are shown in the first~[(a)--(c)], second~[(d)--(f)], and third row~[(g)--(i)]. In the left, middle, and right column the same position on the sample is shown for three consecutive MFM images before, during, and after the collapse of the skyrmion. The position of the collapse is marked by small black arrows. (j)~Frequency shift along the $x$ direction, i.e., line scans, around the skyrmion collapse depicted in panel~(h). The line scan across the collapse (red diamonds) is depicted together with the three scans preceding (gray polygons) and following (gray circles) the collapse.}
	\label{figure4}
\end{figure}

Figure~\ref{figure4} illustrates three such processes, where skyrmions disappear while scanning the image, in detail. Each row consists of three consecutive MFM images of a given sample area recorded before (left column), during (middle column), and after (right column) the collapse of the respective skyrmion. The slow scan direction was from top to bottom, the fast scan direction is denoted by arrows in the top right corner of each panel (from right to left in the top and middle row, from left to right in the bottom row). The skyrmion signal disappears essentially instantaneous in the time scale of the fast scanning and independent of the direction of the latter. We attribute the triggered unwinding to the stray field of the MFM tip either (i)~locally enhancing the magnetic field thereby energetically destabilizing the metastable skyrmion state ultimately triggering its unwinding by means of a Bloch point or (ii)~directly assisting in the creation of a Bloch point due to field gradients. The lack of triggered unwinding processes when external magnetic field and tip magnetization are aligned anti-parallel to each other, see Supporting Information, implies that the former mechanisms dominates. 
As the change of the field is expected to be of the order of ${\sim}15 \ldots 30$~mT,\footnote{In analogy to Ref.~\cite{2006_Engel-Herbert_JApplPhys} and using geometry data and the remanent magnetisation value of 80~emu/cm\textsuperscript{3} given by the manufacturer \cite{Nanosensors} for the SSS-QMFMR tips, we estimate the magnitude of the axial stray field in a distance of 20~nm from the tip apex to be of the order of ${\sim}15 \ldots 30$~mT.} in fact, this finding may represent a direct consequence of the extremely strong field dependency of the skyrmion life time reported from LTEM measurements.\cite{2017_Wild_SciAdv}

In Figure~\ref{figure4}j, as a typical example, the decay of the skyrmion shown in Figure~\ref{figure4}h is further analyzed in terms of single line profiles. Red diamonds indicate the scan during which the unwinding was triggered. For comparison, we also show three line profiles extracted from the same image prior to , i.e., above and after, i.e., below the unwinding, marked by gray polygons and circles, respectively. Prior to the collapse, the line profiles exhibit a clear minimum, characteristic of a scan across the central part of a skyrmion where the local magnetization is anti-parallel to the applied magnetic field. After the collapse, the line profiles essentially correspond to flat lines, characteristic of a topologically trivial conical or field-polarized state. During the collapse, the line profile first tracks the previous line profile for which a typical skyrmion is observed. When coming close the core of the skyrmion, the signal abruptly changes, jumping to a larger value.

\section{Summary}

In summary, the metastable skyrmion state prepared by means of field cooling in the chiral magnet Fe$_{0.5}$Co$_{0.5}$Si was investigated using magnetic force microscopy on the surface of a bulk single crystal. While the metastable skyrmion state is, in principle, of trigonal symmetry, \cite{2010_Munzer_PhysRevB, 2010_Yu_Nature, 2013_Milde_Science, 2017_Wild_SciAdv} it is also subject to sizable disorder. Nevertheless, we find that certain positions are reproducibly decorated by an individual skyrmion after each field cooling process.
Whether skyrmions are formed at these positions under the phase transition at $T_{c}$ or attracted from nearby places under the field cooling in the metastable skyrmion lattice so far remains an open question.
Skyrmions at these positions also turn out to be rather robust against changes of the magnetic field, establishing microscopically that local pinning sites permit an efficient control of position and relative stability of skyrmions in chiral magnets. In addition, we demonstrate that the unwinding of individual skyrmions may be triggered by the MFM tip by means of locally enhancing the applied magnetic field.
Besides reliable nucleation of skyrmions at distinct positions in memory or computation devices, control and fixing of skyrmion positions may be crucial for guiding of spin wave intensity in skyrmion magnonic devices \cite{Schuette_PRB_2014}.
Taken together, our findings suggest that the combination of locally applied magnetic fields and local heat pulses---in the spirit of rapidly quenched skyrmion states reported in Refs.~\cite{2016_Oike_NatPhys, 2017_Nakajima_SciAdv, 2018_Berruto_PhysRevLett}---with the control of skyrmion positions by means of defect pinning may allow to envision single-skyrmion application concepts exploiting materials, in which skyrmions only arise thermodynamically stable in the form of lattices due to bulk Dzyaloshinsky--Moriya interactions.

\section*{acknowledgement}
P.M., E.N., and L.M.E.\ gratefully acknowledge financial support by the German Science Foundation (DFG) through the Collaborative Research Center ``Correlated Magnetism: From Frustration to Topology'' (SFB 1143), the SPP2137 (EN~434/40-1), as well as projects EN~434/38-1 and MI~2004/3-1.
A.B.\ and C.P.\ gratefully acknowledge financial support by the DFG through TRR80 (projects E1 and F2) and SPP2137 (PF393/19-1) as well as by the European Research Council (ERC) through Advanced Grants 291079 (TOPFIT) and 788031 (ExQuiSid).\\
\appendix
\section{Kelvin-probe force microscopy} \label{Appendix:KPFM}
Frequency modulated Kelvin-probe force microscopy (FM-KPFM) allows the determination of the local contact potential difference \cite{Weaver1991,Nonnenmacher1991,Zerweck2005}.
Typically, the Kelvin modulation voltage together with the compensation bias was applied to the sample with an amplitude of 1~V and a frequency of 4.1~kHz, while the tip was grounded.
The resulting sideband in the cantilever motion spectrum is directly detected with a lock-in amplifier within the R9-controller electronics running at the detected cantilever frequency plus the applied modulation frequency. 
While in principle KPFM can be performed simultaneously to MFM measurements, an initial measurement revealed, that no contact potential difference contrast was present on our sample neither in topography nor in MFM mode.
Thus, we fixed the compensation bias in topography mode were the electrostatic interaction is largest and switched off the modulation voltage in MFM mode.
This way, the electrostatic interaction was minimized during the MFM scan without the KPFM controller as additional noise source.
\section{Processing of MFM data} \label{Appendix:Data}
The analysis of the MFM data was carried out by means of batch processing using Python via the Pygwy interface of the Gwyddion analysis software \cite{2012_Necas_CentEurJPhys} in an iterative multi-step process.

In a first step, a preliminary skyrmion map was created by the following algorithm. 
First, the background image measured at high magnetic field was subtracted, yielding a map of the frequency shift $\Delta f$ that is essentially proportional to the out-of-plane stray field of the sample.
This frequency shift map was then filtered five times using a one-point Gaussian filter in order to reduce noise.
Second, in order to find skyrmion positions, as initial guess all pixels with values below 20\% (above 80\%) of the total value span were masked, respectively.
Third, the average frequency shift $\overline{\Delta f}$ and the corresponding root mean square value $\Delta f^{\mathrm{rms}}$ across the unmarked area are computed.
Fourth, new threshold values are defined, namely $\overline{\Delta f} - 2.0\cdot\Delta f^{\mathrm{rms}}$ ($\overline{\Delta f} + 2.0\cdot\Delta f^{\mathrm{rms}}$).
Fifth, again all pixels with values below the lower (above the upper) threshold were masked, respectively.
Sixth, clusters of less than 15 pixels in size were removed from the masks, i.e., areas with an equivalent diameter of $\lesssim 9$~nm.
The remaining clusters were smoothed using a one-point Gaussian filter, subsequently binarizing the mask again by assigning all pixels to a cluster that are associated with values above 20\% of the full scale.
Finally, steps three through six are repeated with the results converging typically in less than ten iterations.

In the resulting preliminary skyrmion map, sometimes skyrmions which are located close to each other were not correctly separated due to the varying signal strength of individual skyrmions.
In principle, this is no problem for the creation of Figure \ref{figure2}(c) and Figure \ref{figure3}(f) but leads to errors in the skyrmion number count.
Thus, we performed in a second step a clusterwise analysis.
For each of the individual clusters in the preliminary skyrmion map we determined the average frequency shift which was then used as threshold value.
Masking all pixels in a grain which lie above the average, leads to a shrinking of the cluster size for clusters containing only one skyrmion whereas for clusters containing several skyrmions the original clusters were split into smaller clusters, i.e., one per skyrmion.
Finally, all clusters were expanded by seven pixels restoring the original size of the detected skyrmion clusters.

Counting separately the survival rate of pinned and non-pinned skyrmions, the algorithm had to assign the property pinned/non-pinned to a detected skyrmion cluster.
Running the detection algorithm only with the initial skyrmion distributions and summing of the resulting skyrmion maps results in Figure \ref{figure2}(c).
Next, the clusters in the inital skyrmion map are sorted into pinned and non-pinned skyrmions with clusters being considered as pinned if they overlap with one of the regions with value nine in Figure \ref{figure2}(c).
Thereafter, we performed the analysis of the decay series.
Multiplying all skyrmion maps of a decay series, leads to the corresponding stability map as shown in Figure \ref{figure3}(e), in which only pixels are marked that are always covered by a cluster in the skyrmion maps.
Adding up the stability maps of the nine cooling cycles results in Figure \ref{figure3}(f).
Finally, for the counting summarized in table \ref{table1}, a pinned/non-pinned skyrmion is considered as survived, if the skyrmion in the initial skyrmion map overlaps with one of the marked clusters in the stability map.

\bibliography{bibliography}
\end{document}